\documentclass{ceurart}

\usepackage{booktabs}
\usepackage{graphicx}
\usepackage{amsmath}
\usepackage{amssymb}
\usepackage{multirow}
\usepackage{array}
\usepackage{tabularx}

\begin{document}

%% Rights management information (CC-BY is the CEUR-WS default).
\copyrightyear{2026}
\copyrightclause{Copyright for this paper by its authors.
  Use permitted under Creative Commons License Attribution 4.0
  International (CC BY 4.0).}

%% Conference information.
\conference{GenAIECommerce'26: The Third Workshop on Agentic and Generative AI
  for E-Commerce, co-located with RecSys, September 28, 2026, Minneapolis, MN, USA}

\title{AURA: Agentic Diagnosis and Refinement for Production Recommender Systems at Scale}

\author[1]{SungGeun Kim}[email=snugkim@gmail.com]
\fnmark[1]
\fnmark[2]
\author[1]{Abhinav Narain}[email=abhinav.narain@disney.com]
\fnmark[1]
\author[1]{Daniel Nemirovsky}[email=daniel.nemirovsky@disney.com]
\fnmark[1]
\address[1]{The Walt Disney Company, San Francisco, CA, USA}
\fntext[1]{All authors contributed equally to this research.}
\fntext[2]{Work done while at The Walt Disney Company. Now at Intuit.}

\begin{abstract}
How and why does a recommender system fail the users it serves? Oftentimes in recommender systems, 
practitioners are left to improve their algorithms based
on a combination of feedback from stakeholder teams, domain expertise, and insights from data analyses.
Yet, the nuances of how and where the recommendations are performing well or poorly for
the end users are difficult to discern from aggregate quantitative metrics. 

Whereas these metrics can provide a high-level and incomplete picture, further
granularity into the quality of recommendations and their patterns requires reasoning
with domain understanding and objectivity, at scale. In this paper, we contemplate this complex conundrum
and describe a method and implementation that utilizes the latest AI agentic advances to provide
actionable diagnoses and improvements for the production recommender systems. 

We present AURA (Agentic Understanding and Refinement of recommender Algorithms), an
end-to-end agentic system that performs qualitative evaluation at scale and can then generate 
improvements to our algorithms at the code level. Specialized
agents read production engagement logs, from thousands of sessions to millions, and
surface patterns and examples of how the recommender fails real users. The next step uses those diagnoses
as well as context about the recommender's own code,
data, and training pipeline to propose and implement refinements grounded in that codebase. We report the
system design, initial tests on production data from two large consumer platforms
at a major media-streaming company, safeguards, operational learnings, and early results toward a self-improving
recommender system. Finally, the diagnostic gap AURA closes is not specific to streaming. The
architecture is built to transfer and every domain-specific element enters through the
configuration layer that already ported it between our two platforms. We map it
concretely to e-commerce and online-retail recommendation.
\end{abstract}


\maketitle

\section{Introduction}

Aggregate evaluation metrics (AUC, NDCG, precision/recall, diversity, coverage) are the
standard quantitative toolkit for recommender systems~\cite{gunawardana2015evaluating,
zangerle2022evaluating}. They tell teams whether a model is, on average, doing better or
worse than a baseline. However, these metrics may obfuscate finer grained qualitative issues 
occurring in certain situations for items and/or users. For instance, even though a model can improve
across aggregate metrics, it may quietly be surfacing undesirable or egregious recommendations, introducing diversity collapses,
personalization regressions, or temporal staleness. McNee et al.\ argued that being
accurate is not enough and accuracy-centric evaluation misses what makes recommendations
useful to users~\cite{mcnee2006accurate}.
More recently, Edizel et al.\ reported large offline improvements at Netflix that did not
translate to online gains, because content-type effects do not show up in aggregate
scores~\cite{edizel2024gaps}. Broad aggregation can induce Simpson's
paradox~\cite{jadidinejad2021simpsons, prost2022simpsons} and mask severe degradation for
specific demographic subgroups, tail items, or content
producers~\cite{ekstrand2018coolkids, mehrotra2018fair, mitchell2019modelcards,
shani2011evaluating, yao2017parity}. 

Finding the correct slices and segments for deeper eval analyses, or augmenting the eval metric set,
requires acute domain knowledge and engineering effort, especially as these scopes
evolve.  This diagnostic gap is not specific to any one domain. An e-commerce product feed that
quietly over-promotes high-margin items to price-sensitive shoppers, or a retail search
ranker that buries long-tail inventory for a subset of queries, exhibits the same
structural failure: segment-level harm averaged away by a healthy top-line conversion or
relevance score.

For industry practitioners this is an operational bottleneck. When a metric regresses
offline or online, the cause could be a regional relevance mismatch, a feedback loop
narrowing exposure, a deduplication failure across shelves, a cold-start spike, or
upstream data corruption. For example, an engineer working on a regression on Platform~B can
read that NDCG dropped and still have no idea whether the issue is kids' profiles getting
horror promos, sports fans getting reruns, or a deduplication bug across the homepage
shelves. Top-line metrics tell us that something is wrong but they do not tell us why.
A/B tests, the standard tool for causal evaluation, share the same diagnostic blind spot,
and they are expensive, slow, and risky on high-impact surfaces~\cite{kohavi2020trustworthy}.
Data scientists and ML engineers are tasked with maintaining and improving the recommender systems often measured
by these aggregate quantitative metrics. Yet, without an understanding of the core issues the 
users are facing regarding their recommendations, the engineers are left with little insight into 
where the actual areas of concern and opportunity are. To account for this, practitioners typically
complement quantitative metrics with manual, and hence small scale, qualitative analyses including
slicing cohorts, writing ad-hoc queries, and watching individual sessions. It is this
latter type of analysis, detecting issues and patterns from the engagement history and
context using domain understanding, objectivity, and reasoning, that the rise of AI
agentic systems newly enables performing at scale.

In this paper, we present AURA, an agentic system that automates the workflow from
qualitative evaluation through diagnosis to resolution at production scale. Specialized agents read production sessions, can call tools to provide informative platform, user, and content context, and report how the
system benefits and fails real users, with severity and evidence attached. They then read
the recommender's own code, data, and training pipeline and propose refinements grounded
in that codebase. We have implemented an initial version and run it on production
engagement data from Platform~A and Platform~B at a major media-streaming company,
producing diagnostics, technical proposals, and pull requests for our ML engineers to
review. An engineer stays in the loop and the pipeline can run iteratively and learn from its
prior experiments using a memory log.
This work provides the following contributions:
\begin{enumerate}
  \item \textbf{An end-to-end agentic design for diagnosing and refining production
  recommender systems.} To our knowledge, AURA is the first system to place an
  evidence-backed failure taxonomy between agentic evaluation at scale and code-level
  refinement. It derives the taxonomy from production session evidence, then works each
  finding back into the recommender's own codebase. We detail
  the design and implementation of AURA, a multi-stage agentic pipeline that selects
  sessions from production engagement data, diagnoses how the recommender benefits and
  fails users, generates root-cause hypotheses grounded in the recommender's own code and
  training pipeline, and produces technical proposals for resolution. The architecture is
  extensible to autonomous implementation, training, and evaluation across different domains.
  \item \textbf{Implementation and production deployment at Platform~A and Platform~B.}
  We document the production implementation behind the design. An
  architecture-versus-prompt attribution methodology separates structural pipeline gains
  from prompt-edit gains across iterative refinement passes, lifting stage-level rubric
  scores from 15/25 to 23/25 on Platform~A and from 17/25 to 24/25 on Platform~B
  (\S\ref{sec:diagnostic-quality}). A cost-efficient multi-model orchestration sends the
  high-volume classification work to lightweight models and reserves frontier models for
  interpretation. Upstream per-session judging dominates end-to-end cost and scales with
  population (\$321 over 96,801 Platform~A sessions, \$250 over 101,594 Platform~B
  sessions), while AURA's downstream aggregate analysis adds a small fraction on top
  (\S\ref{sec:cost}). A human-in-the-loop workflow promotes findings to action and
  dismisses what does not survive scrutiny (\S\ref{sec:safeguards}).
  \item \textbf{Operational learnings from running AURA on production data.} After the
  initial discovery phase, locking the category vocabulary clearly outperformed
  open-ended classification. Post-hoc cleanup shrank to a single rename with zero merges
  on Platform~B, and both platforms converged on closed taxonomies, with consolidation
  collapsing Platform~A's 12 emergent tags to 8 categories with no diagnostic loss
  (\S\ref{sec:diagnostic-quality}). Engineer trust is built through structured evidence
  rather than narrative explanations (\S\ref{sec:safeguards}).
  \item \textbf{Safeguards and accountability for agents touching production data and
  code.} Every agent runs read-only against analytical data lakes, code changes happen in
  a sandboxed clone, promotion requires a named engineer, and dismissal is a first-class
  outcome (\S\ref{sec:safeguards}). We catalog the defensive-validation layer this
  demanded in production, from fabricated-identifier sanitization to
  failure-taxonomy collapse. (\S\ref{sec:lessons}).
\end{enumerate}

\section{Related Work}

The LLM-as-Judge paradigm~\cite{zheng2023judging} and agentic
evaluators~\cite{zhang2025nohuman} can evaluate recommendations without
task-specific training. LLMs have enough competence for zero-shot ranking~\cite{hou2024zeroshot}. 
However, judges exhibit persistent biases including
presentation order sensitivity~\cite{wang2023fair}, ``style over substance''
preferences~\cite{wu2023style}, and unwarranted assertiveness~\cite{hosking2024feedback}.
Dedicated autoraters like FLAMe~\cite{vu2024flame} improve reliability but require
task-specific training, and all such systems ultimately require ``human grounding'' for
trustworthy judgments~\cite{krumdick2025nofree}. Closest to our evaluation layer, Zhang
et al.~\cite{zhang2025nohuman} remove humans from the loop to scale agentic evaluation of
recommendations. AURA consumes that kind of per-session verdict as raw material rather
than as the end product, aggregating verdicts into a failure taxonomy and following each
finding into the recommender's own code. Because LLMs can interpret rich session
context but remain sensitive to prompt framing and aggregation choices at scale, AURA
uses them inside a staged diagnostic workflow with filtering, structured outputs, and
verification rather than treating any single LLM judgment as final. Programmatic
evaluation tools such as Slice Finder~\cite{chung2019slicefinder} and
CheckList~\cite{ribeiro2020checklist, chia2022reclist} expose slice-specific or
behavior-specific failures, but they do not produce root-cause hypotheses from
recommendation-session evidence and domain context. AURA uses LLMs to generate it and to
explain recurring failure categories after they have been surfaced.

A separate line evaluates recommenders by simulating users with LLM agents. These include generative
user agents in Agent4Rec~\cite{zhang2024agent4rec} and RecAgent~\cite{wang2023recagent}. Additionally,
iEvaLM~\cite{wang2023ievalm} uses LLM-based user simulators for conversational
recommendation, and the broader RecAI toolkit~\cite{lian2024recai} packages
similar evaluation and explanation tooling. Whereas simulation provides some
counterfactual control, AURA sees the real production sessions
as the higher-fidelity evidence, and diagnoses from those.

Wang et al.~\cite{wang2026selfevolving} demonstrate that LLM agents can autonomously
optimize recommendation models at YouTube scale through code generation and A/B testing
($O(100)$ experiments/week). Kim et al.~\cite{kim2026selfevolverec} propose ``directional
feedback'' co-evolution, though validation remains limited to offline datasets rather
than live production environments. Broader agentic-ML benchmarks frame the wider
landscape of LLM agents doing machine-learning experimentation and
engineering~\cite{huang2023mlagentbench, chan2024mlebench, jiang2025aide}. These systems
search over model space (architecture, loss formulation, training schedule,
hyperparameters) and optimize against aggregate engagement signals. Critically, AURA starts from the
other end. It first diagnoses user-facing failures from recommendation sessions, then
grounds candidate refinements in the recommender's own code and training pipeline.
The closed-loop extensions we sketch as future work would meet these systems in the
middle.

\section{System Architecture}
\label{sec:architecture}

\subsection{Overview}

\begin{figure}
  \centering
  \includegraphics[width=\linewidth]{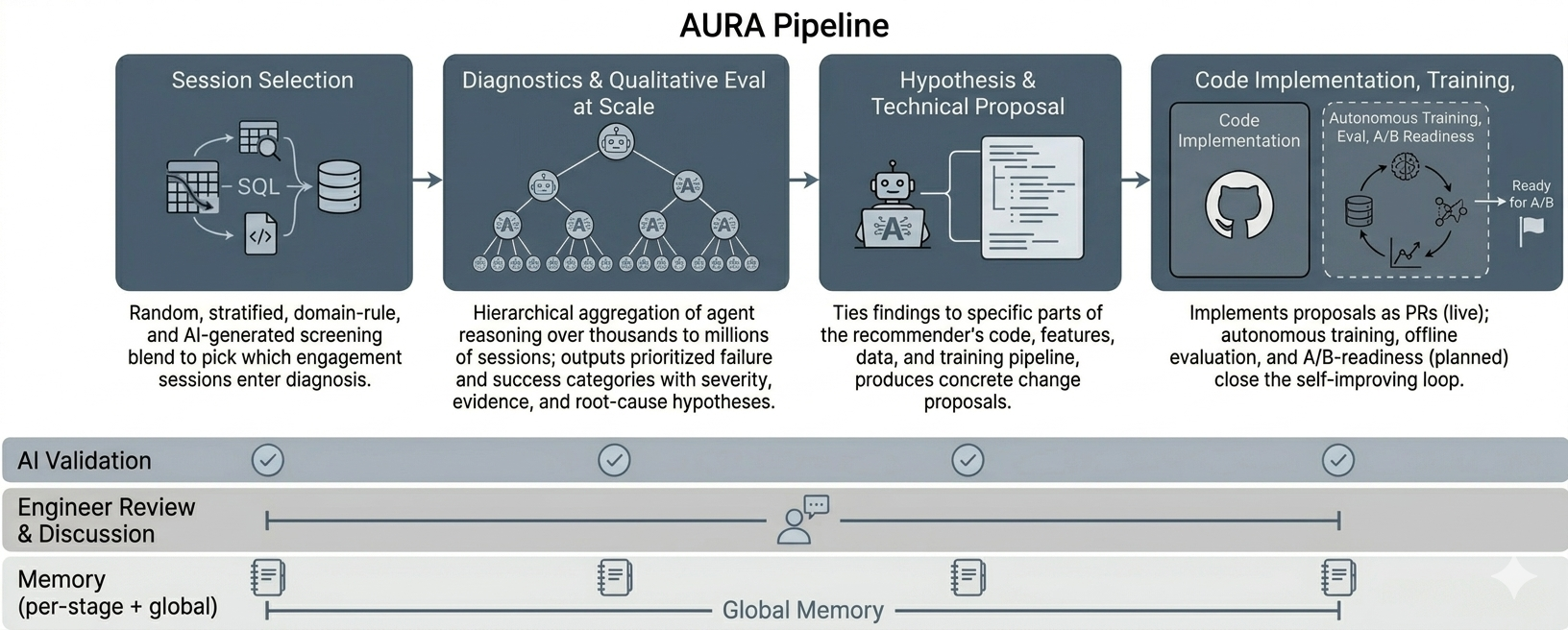}
  \caption{The AURA pipeline: Session Selection; Diagnostics \& Qualitative Eval at
  Scale; Hypothesis \& Technical Proposal; and Code Implementation, Training, Eval, A/B
  Readiness, with AI validation, engineer review, and memory running across all four
  stages. The dashed portion of the final stage is a planned extension. The current
  deployment delivers it as far as engineer-reviewed pull requests.}
  \label{fig:pipeline}
\end{figure}

AURA is an agentic pipeline that consumes production session logs and emits evidence-backed
change proposals against the recommender's own code. The workflow runs as four sequential
stages, with AI validation, engineer review, and memory threaded through all of them
(Figure~\ref{fig:pipeline}).

A selected stage chooses which sessions to examine. Selected sessions enter a diagnostic stage (\S\ref{sec:diagnostic-stage}) that scales LLM reasoning over large session volumes through hierarchical agent aggregation. The
diagnostic stage applies configurable quality criteria to judge how recommender
served users to find problematic patterns. It then emits failure and success categories, each with a severity, supporting evidence, and a root-cause hypothesis. These feed a hypothesis-and-proposal stage
(\S\ref{sec:hypothesis-stage}) that reasons over the recommender's own code, features,
and training pipeline. The technical proposals it produces then feed to a code-implementation
stage (\S\ref{sec:code-stage}) that turns them into pull requests today, and (in the
planned extensions) into training, offline evaluation, and A/B-ready artifacts that close
the self-improving loop. Agents are given tools to gather additional information
about the platform, users, items, and engagement where appropriate and available.

The system is designed for flexibility in that each stage can permit alternative
instantiations. For example, session selection can blend random, stratified, domain-rule, and
AI-driven mechanisms. The diagnostic aggregation can use a flat panel or a hierarchical tree mechanism and the
hypothesis generation can read a narrow slice of the codebase or the full repository. Lastly, the
code-change proposals can stop at natural-language descriptions or push to executable
patches. A new platform is a configuration change, not a code change. The data sources,
column mappings, segment definitions, and prompt overrides are supplied per platform, and
platform isolation is enforced by a required parameter that threads through every
component. Cost is managed by deterministic orchestrator by routing sessions at each stage to
a model suited to its complexity, sending high-volume work (classification,
consolidation) to lightweight models and reserving frontier models for analysis,
verification, and code generation. We report concrete cost figures in \S\ref{sec:cost}.

\subsection{Session Selection Stage}
\label{sec:selection-stage}

AURA's first stage selects which sessions enter deep diagnosis (Figure~\ref{fig:pipeline}). This mechanism is
configurable, blending random or stratified sampling with engineer-written domain SQL and
AI agents. A profile pass summarizes the data distribution for a shared numerical
baseline, then samples the segments whose statistics flag issues. The domain SQL catalog draws on recurring pain points from Platform~A/B production experience (low NDCG, high watch time, repeat-content counts,
sub-genre and content-tone mismatch, kid-profile safety) and research-literature failure
modes (popularity bias, position bias, distribution shift, diversity collapse). Multiple
independent LLM-powered screening agents then use tools to compose queries that identify
which sessions to analyze.

The flagged segments go to $N$ agents that independently propose candidate problem
cohorts in parallel. A dedup pass merges overlapping proposals and consensus step ranks the survivors by majority vote
across agent proposals. A surviving cohort might be, for
example, kids’ profiles in one region receiving horror promotions. Each surviving
cohort’s sessions are decorated with light per-session context: the judge verdict and reason, enrichment columns
(editorial flags, user history, item timestamps, impression actions), and aggregated
summaries for the next stage.

\subsection{Diagnostic Stage}
\label{sec:diagnostic-stage}

The diagnostic stage turns selected sessions into a prioritized set of failure and
success categories, each with a severity, supporting evidence, a root-cause hypothesis,
and a GOOD/BAD verdict with free-text reasons (\S\ref{sec:diagnostic-quality}). Verdicts
follow an LLM-judge paradigm under configurable, extensible quality criteria.

The challenge is scale since production runs cover hundreds of thousands to millions of
sessions, far beyond what any single agent can hold in context. The design solves this
through hierarchical aggregation. At the leaves, parallel agents each consume a tractable
slice (on the order of a thousand sessions, within context) under category-specific
prompts that surface emergent good and bad patterns with evidence. Findings propagate to
peer agents that reason about overlap and significance, consolidate near-duplicates, and
surface a smaller, higher-confidence list. This repeats up the tree until a single layer
produces the final prioritized list.

The current Platform~A and Platform~B deployment instantiates this hierarchy as a
six-step funnel. \emph{Classification} runs in parallel batches, classifying each session
against an emergent or closed taxonomy and assigning a session type and short
description. \emph{Consolidation} is the first cross-batch aggregation. It is a single LLM
pass that merges near-duplicate categories, renames noisy labels, and drops low-signal
entries (in one Platform~A run, ``duplicate items across shelves'' and ``repetitive
recommendations'' merged into ``cross-shelf deduplication failure''). \emph{Sampling}
retains only categories worth a frontier-model deep dive (via a
top-$N$/count/percentage rule, e.g., the largest categories, at least roughly a hundred
sessions, or five percent of the type's total) and draws a bounded random sample of
sessions per retained category, while safety-critical categories (e.g., content/policy
violations) bypass the threshold. \emph{Analysis} runs in parallel across retained
categories, one analyst worker per category producing a structured finding (description,
severity, root-cause hypothesis, supporting evidence, actionable recommendations).
\emph{Verification} sits over the findings. One agent cross-checks consistency, validates
claims against the sampled sessions, and corrects severity miscalibrations.
\emph{Synthesis} produces an executive summary that groups failures and successes,
surfaces cross-cutting observations, and outputs prioritized recommendations.

\subsection{Hypothesis and Technical Proposal Stage}
\label{sec:hypothesis-stage}

Diagnostic findings highlight what is going wrong while the next stage proposes what
to change and why. For each verified finding, the hypothesis-and-proposal stage pulls in
the recommender's own context such as model code, feature definitions, training pipeline,
label specifications, as well as data statistics and schemas, and reasons about which parts of the system could
be producing the finding. The output is a structured technical proposal that includes a hypothesis
tying the finding to specific components (for example, ``the ranking model over-weights
global popularity signals in the cold-start branch'') and one or more candidate change
suggestions targeted at those components.

We use the term ``root-cause hypothesis'' deliberately. AURA does not deliver formal causal
inference. It produces explanatory hypotheses grounded in code, data,
and training context. The surrounding evidence (sampled session excerpts, the diagnostic
agent's reasoning, and schema and code references) is what makes those hypotheses
tractable for engineers to verify before committing to a change.

\subsection{Code Implementation, Training, Evaluation, and A/B Readiness Stage}
\label{sec:code-stage}

The final stage takes technical proposals through to code. A generate-evaluate-refine
loop~\cite{madaan2023selfrefine} prompts a frontier model with the verified finding, the
root-cause hypothesis, schema metadata for the relevant tables, and a code index of file
paths and line counts (rather than full source, which would exceed context limits). A
separate evaluator model scores each suggestion for correctness, feasibility, and side
effects and returns a pass/fail verdict with feedback. Failures, cycle back to the
generator. A programmatic pass then checks every surviving suggestion against the actual
codebase for nonexistent files, missing imports, or fabricated references.

The current deployment delivers this stage as far as the pull request. Surviving
suggestions are written to the suggestions database and surfaced through the shared
review interface (\S\ref{sec:validation}). Two extensions under active development
would close the loop end-to-end. The first
triggers an automated training run when a promoted suggestion is merged into the
recommender's sandboxed repository, and the second runs offline evaluation on the trained
artifact and packages an A/B-ready result. With both in place, AURA iterates autonomously
between engineer-review checkpoints, proposing, implementing, evaluating, and
refining until the offline evaluation clears.

\subsection{Validation, Engineer Review, and Memory}
\label{sec:validation}

Validation, engineer review, and memory run through every stage.
Validation means that every LLM output boundary carries a programmatic check.
Session-ID references are checked to exist in the input, and JSON-schema validation with
a regex fallback catches malformed payloads.
The diagnostic stage adds a
verification agent that cross-checks consistency across findings. The code-implementation
stage adds a validator confirming file paths and imports resolve against the actual
repository. Parse failures and violations are logged to a structured data-quality flags
array and surfaced to engineers.

Engineers, for their part, may interact with every stage through a shared web interface that
pairs raw evidence with AURA's structured outputs. They can read sampled sessions,
discuss findings with the relevant agents, vote, comment, promote items into team
backlogs, or override AURA's calls. Human review is a property of the pipeline, not
a final-stage gate.

Memory is a key consideration that closes the loop across runs. Each stage logs the categories
surfaced, hypotheses tried, code changes proposed, and outcomes observed, which a shared
memory layer aggregates across stages and runs. With this memory, each run builds on what
earlier runs found and what already failed, so AURA moves forward instead of repeating
itself.

\section{Evaluation}
\label{sec:evaluation}

We considered three main questions which drive our evaluation. Are the findings real? What does a diagnostic run
cost? And does the same pipeline produce actionable findings on a platform it was never
tuned for?

\subsection{Experimental Setup}
\label{sec:setup}

\paragraph{Data.} We evaluate on production recommendation sessions from two streaming
platforms within a large media enterprise: Platform~A (family-oriented) and Platform~B
(general-audience). Platform~A contributed 18,901 sessions flagged as BAD by the upstream
per-session judge out of 96,801 total sessions evaluated (about 19.5\%). Platform~B
contributed 4,154 BAD sessions out of 101,594 evaluated (about 4.1\%). We do not read the
19.5\% versus 4.1\% gap as a quality ranking between platforms. Rather, judges are prompted and
calibrated per platform, and Platform~B's prompt was revised mid-stream, so the rates are
not directly comparable. These represent
all negatively-evaluated sessions from production recommendation serving over the
evaluation window, not a sample. Sessions span diverse user cohorts, regions, devices,
and content catalogs.

Because AURA's diagnostic quality depends on the upstream per-session judge, we
separately validated the judge following the LLM-as-judge framework~\cite{zheng2023judging,
verga2024juries}, using a panel of three independent evaluators (Claude Opus 4.6,
GPT-5.5, and Gemini 3.1 Pro) with majority-vote aggregation. On a stratified validation
sample of $\sim$200 sessions on both platforms, the panel assessed whether
each judge verdict was correct, flagging error patterns such as demographic overfit,
position bias, and overconfidence. Majority-vote agreement (2+ evaluators judging the
verdict correct) was 96.0\% on Platform~A (192/200) and 87.6\% on Platform~B (176/201),
with full three-way agreement at 80\% and 58\% respectively. We treat the panel as a
calibration check, not hard ground truth. At the high positive prevalence here (near 0.9)
neither raw agreement nor chance-corrected agreement is decisive, and
diagnostic-correctness is a genuinely hard, partly subjective
judgment.\footnote{Chance-corrected agreement is low by construction at this prevalence
(Fleiss' $\kappa$ $\approx$ 0.17--0.38 across platforms and label granularities,
``slight'' to ``fair''): when evaluators almost always agree, chance agreement is high and
the residual is necessarily small. The informative figure is that a majority explicitly
disagrees with the judge on only 3--4\% of the validation sample, with non-trivial
``unsure'' rates (18\% on Platform~A, 36\% on Platform~B) that $\kappa$ captures and raw
agreement does not. One caveat: Gemini-family models appear both upstream and downstream,
which can inflate agreement through self-preference~\cite{panickssery2024selfpref}.}

\begin{table}
  \caption{Per-stage rubric evaluation results: baseline $\rightarrow$ after iterative
  improvement. Fractions indicate GOOD criteria out of total criteria for each stage.}
  \label{tab:rubric-results}
  \begin{tabular}{lcccc}
    \toprule
    & \multicolumn{2}{c}{Platform A} & \multicolumn{2}{c}{Platform B} \\
    Pipeline Stage & Before & After & Before & After \\
    \midrule
    Classification    & 4/5   & 5/5   & 4/5   & 5/5 \\
    Consolidation     & 1/4   & 3/4   & 2/4   & 4/4 \\
    Category analysis & 3/5   & 5/5   & 4/5   & 5/5 \\
    Verification      & 2/3   & 3/3   & 2/3   & 3/3 \\
    Synthesis         & 4/4   & 4/4   & 4/4   & 4/4 \\
    Code suggestions  & 1/4   & 3/4   & 1/4   & 3/4 \\
    \midrule
    Total             & 15/25 & 23/25 & 17/25 & 24/25 \\
    \bottomrule
  \end{tabular}
\end{table}

We evaluate diagnostic quality through binary rubrics applied independently to each
pipeline stage's output, since systemic recommendation failures lack a single definitive
root cause and standard classification benchmarks do not apply. For each of the six
scored stages we define 3 to 5 criteria (Table~\ref{tab:rubric-criteria}), each testing a
single observable dimension with an unambiguous GOOD/BAD threshold. These six scored
stages contribute 25 criteria (score is the count of criteria rated GOOD) in total and are the rubric's unit of evaluation (Classification, Consolidation, Category analysis, Verification, Synthesis, and Code suggestions). They differ from the diagnostic funnel (\S\ref{sec:diagnostic-stage}) in only two places: the funnel's Sampling step is not scored here, and the rubric adds Code suggestions from the code-implementation stage (\S\ref{sec:code-stage}). This per-stage, per-criterion evaluation mirrors the behavioral testing methodology of CheckList~\cite{ribeiro2020checklist} and the
recsys-specific analog RecList~\cite{chia2022reclist}, which measure failure rates per
capability rather than a single aggregate accuracy. Each criterion is evaluated by two
independent LLM judges (Gemini 3.1 Pro and GPT-5.5) examining the pipeline's actual
output, and inter-judge agreement determines the final verdict.

For each pipeline stage we follow a systematic improvement cycle: (1) evaluate the
baseline output against rubrics for both platforms, (2) identify BAD criteria and their
root causes, (3) apply targeted changes (architecture fix or prompt revision), (4)
re-evaluate against the same rubrics. We keep a change when more criteria
flip (BAD$\rightarrow$GOOD) than GOOD$\rightarrow$BAD. A criterion flipping the
other way blocks the change, unless the two-judge disagreement finds a stricter
judge rather than a worse output. This protocol separates architecture-level
fixes (code changes affecting data flow) from prompt-level fixes (template changes
affecting LLM instructions), so quality improvements can be attributed to their actual
cause.

\subsection{Diagnostic Quality}
\label{sec:diagnostic-quality}

\paragraph{Rubric-based evaluation results.} Table~\ref{tab:rubric-results} reports the
per-stage results on both platforms before and after the full improvement cycle. Six
architecture fixes plus targeted prompt revisions moved both platforms to near-full GOOD
across the 25 criteria of Table~\ref{tab:rubric-criteria}.

\paragraph{Architecture vs.\ prompt attribution.} Most quality improvements came from
architecture-level fixes, not prompt engineering. Of six architecture changes
(display-name normalization, threshold tuning, adaptive sample sizes,
response-truncation limits, classification-context forwarding, and specialized verifier
prompts), three resolved rubric failures that no prompt change could fix. For instance,
the severity-data alignment failure in category analysis was not the LLM ignoring
severity rules. The prompt template simply never received the category's session
percentage, and adding it resolved the issue on both platforms simultaneously. Only two
stages (classification and category analysis) needed prompt-level changes. The other four
reached full GOOD through architecture fixes alone or needed none.

\paragraph{Closed taxonomy as a classification technique.} Our most impactful
prompt-level finding was that closed taxonomies (explicit category enumeration with a
``DO NOT invent new categories'' instruction) clearly outperform open-ended
classification in production. On Platform~B, the open-ended baseline produced 30+
overlapping categories. Adding format and consolidation rules cut this to 22 unique tags
but did not eliminate proliferation. The final closed taxonomy (13 defined categories)
reached 5/5 GOOD and cut post-hoc cleanup to almost nothing. Classification emitted 16 tags (the model
occasionally falls back to off-taxonomy labels) and consolidation reduced to a single
rename with no merges or deletions, versus the multi-merge cleanup the baseline required.
Platform~A converged independently on the same approach (a 13-entry enumeration with the
same instruction). Its run produced 12 tags that consolidation collapsed to 8 categories
(one merge of three synonymous tags, five renames, no deletions). The transferable
insight is that database-populated category vocabularies from prior runs can be
overridden by explicit prompt instructions, which requires careful coordination between
the prompt template and the vocabulary enrichment system.

\begin{table}
  \caption{Diagnostic output from a representative production run on each platform:
  failure categories with severity, session count, and share of the BAD population.}
  \label{tab:diagnostic-output}
  \begin{tabular}{llrr}
    \toprule
    Severity & Category & Sessions & \% \\
    \midrule
    \multicolumn{4}{l}{\emph{Platform A}} \\
    Critical & Genre Pref.\ Mismatch & 9,135 & 48.3 \\
    Critical & Sub-Genre Pref.\ Mismatch & 6,444 & 34.1 \\
    High & Franchise Pref.\ Mismatch & 1,571 & 8.3 \\
    High & Content Type Mismatch & 1,036 & 5.5 \\
    High & Regional Pref.\ Mismatch & 978 & 5.2 \\
    Medium & Age-Inappropriate Suggestions$^\dagger$ & 533 & 2.8 \\
    Low & Evaluator Label Mismatch$^\ddagger$ & 250 & 1.3 \\
    Low & User-Age Content Era Mismatch & 163 & 0.9 \\
    \midrule
    \multicolumn{4}{l}{\emph{Platform B}} \\
    Critical & Genre Pref.\ Mismatch & 1,911 & 46.0 \\
    Medium & Tone \& Sensibility Mismatch & 310 & 7.5 \\
    Low & Content Format Mismatch & 93 & 2.2 \\
    Low & Age-Inappropriate Suggestions$^\dagger$ & 76 & 1.8 \\
    Low & Franchise Affinity Mismatch & 25 & 0.6 \\
    \bottomrule
    \multicolumn{4}{l}{\footnotesize $^\dagger$Safety-relevant; retained via top-$N$/count rule.} \\
    \multicolumn{4}{l}{\footnotesize $^\ddagger$Meta-category: pipeline flagged the upstream judge's own verdict as likely incorrect.} \\
    \multicolumn{4}{p{0.9\linewidth}}{\footnotesize Severity is threshold-derived from
    session \% (Platform A: Critical $\geq$10\%, High 5--10\%, Medium 2--5\%, Low
    $<$2\%; Platform B: Critical $\geq$25\%, High 10--24\%, Med 3--9\%, Low $<$3\%).
    Platform A categories are non-exclusive (a session may carry multiple tags), so
    percentages sum to $>$100\%; 17,996 of 18,901 BAD sessions (95.2\%) were categorized.
    Platform B's higher thresholds retain fewer categories: 2,415 of 4,154 (58.1\%) fall
    in a surfaced category. The rest fell below threshold or were non-diagnostic.} \\
  \end{tabular}
\end{table}

\paragraph{Diagnostic output quality.} Table~\ref{tab:diagnostic-output} shows the final
diagnostic output from a representative production run on each platform. The pipeline
produces a severity-stratified taxonomy in which categories are ordered by session share
via configurable per-platform thresholds (Table~\ref{tab:diagnostic-output}). Severity
here is a prevalence tier. It says how many sessions a failure touches, not how much harm
each instance does. Frequency and harm are different axes, which is why safety-relevant
categories carry a retention override (\S\ref{sec:diagnostic-stage}) and surface on both
platforms despite low counts. Both
platforms independently identify genre preference mismatch as a dominant failure mode,
but the manifestations differ (sub-genre and franchise affinity for Platform~A's family
catalog vs.\ tone-and-sensibility mismatches on Platform~B's general audience),
suggesting the pipeline discovers platform-specific patterns rather than generic
complaints. The Platform~A run also turned on its own judge. In an Evaluator Label
Mismatch category (250 sessions), the analysis stage, examining the full session
evidence, concluded that the upstream per-session verdict was itself incorrect. Instead of
inheriting the judge's mistakes silently, the pipeline caught the judge being wrong.

The two runs also surfaced different root causes: Platform~A identified a systemic
popularity bias, where globally trending content dominated across all failure categories,
traced to the ranking model over-weighting popularity signals; Platform~B, a
content-sensitivity gap, where insufficient demographic filtering surfaced
age-inappropriate recommendations misaligned with user profiles (its Age-Inappropriate
Suggestions category).

\paragraph{Individual stage contributions.} Several of these categories survive only
because of the retention rule. Changing it from a percentage-only threshold to a
top-$N$/count/percentage threshold surfaced Platform~A's low-frequency categories (e.g.,
User-Age Content Era Mismatch, 163 sessions, 0.9\%) and all five of Platform~B's failure
categories; a percentage-only ($\geq$5\%) rule would have kept only Platform~B's two
largest and discarded the three smaller but still-actionable ones (Content Format,
Age-Inappropriate Suggestions, Franchise Affinity). Downstream, the verification agent
cross-checked severity calibration across categories, and the cross-stage validation
layer (\S\ref{sec:validation}) stripped hallucinated session identifiers before findings
reached engineers.

\paragraph{Closed-loop fix validation.}
\label{sec:closed-loop}
The strongest test of whether a finding is real is whether a fix it motivates survives
scrutiny. We screened AURA's candidate fixes offline, against the cohort they were meant
to fix, before any could spend a production A/B slot. For the dominant Genre Preference
Mismatch category, the remediation stage produced two plausible changes: (1) an explicit
user-genre $\times$ candidate-genre cross feature, and (2) a candidate-aware attention
mechanism pooling a user's genre-watch history conditioned on the candidate. Yet neither
improved ranking on the diagnosed cohort (1,911 Platform~B sessions), and both tracked the
production baseline to within run-to-run noise on aggregate metrics ($\pm$0.1\%,
Table~\ref{tab:offline-metrics}). Offline screening thus leaves the fix an open candidate:
safe to iterate on, not yet demonstrated.

\begin{table}
  \caption{Global offline metrics: candidate-aware fix vs.\ the production-clone
  baseline, as a percentage delta on each of three independently-trained days. Every
  delta is within $\pm$0.1\% with no day regressing beyond noise.}
  \label{tab:offline-metrics}
  \begin{tabular}{lrrr}
    \toprule
    Metric (global) $\Delta$ & Day 1 & Day 2 & Day 3 \\
    \midrule
    Click AUC & +0.03\% & +0.06\% & +0.01\% \\
    WM AUC & +0.04\% & +0.08\% & +0.00\% \\
    WM weighted AUC & +0.10\% & +0.06\% & +0.04\% \\
    NDCG@10 & $-$0.01\% & +0.05\% & +0.02\% \\
    \bottomrule
  \end{tabular}
\end{table}

\subsection{Cost-Effectiveness}
\label{sec:cost}

AURA's diagnosis adds little to the cost of the per-session judging it builds on. Across
both platforms, upstream judges account for the overwhelming majority of spend
(92--99\%), while AURA's aggregate analysis over BAD-verdict sessions is about 8\% of
the total on Platform~A and about 1\% on Platform~B (Table~\ref{tab:cost}). Multi-model routing is what
keeps that share small: Gemini~3 Flash handles high-volume classification and
consolidation, and the more expensive Gemini~3.1 Pro is reserved for the low-volume
category analysis, verification, and synthesis stages.

Within AURA's own stages, classification consumes the majority of tokens (41.8M on
Platform~A vs.\ 369K on Platform~B), and the rest operate on progressively smaller
category-level aggregations. Platform~A's higher aggregate cost is driven by larger,
higher-context classification batches (757 calls at $\approx$55K input tokens vs.\
Platform~B's 41 at $\approx$9K), which prompt caching partly offsets.

In absolute terms, aggregate diagnosis costs at most \$5.38 per actionable finding on
either platform, and \$43--\$50 per finding end-to-end once upstream judging is included.
Runtime follows the same split: of the 1,554 and 965 minutes for an end-to-end run on
Platform~A and Platform~B, AURA's own pipeline accounts for only 304 and 72 minutes. Because the
end-to-end total is set by per-session judging, which scales linearly with the session
population, the marginal cost of adding AURA to an existing evaluation stack is small.

\begin{table}
  \caption{Rubric criteria per pipeline stage. Each criterion is binary (GOOD/BAD) with
  an unambiguous threshold observable from the stage's output.}
  \label{tab:rubric-criteria}
  \begin{tabular}{lcp{0.55\linewidth}}
    \toprule
    Pipeline Stage & Number of Criteria & Description \\
    \midrule
    Classification & 5 & No catch-all categories; each category maps to one distinct
    fix; full session coverage; taxonomy compliance; descriptions contain specific
    pattern + code area \\
    Consolidation & 4 & Merges combine genuinely synonymous categories; distinct
    failures remain separate; renames are more specific than originals; valid output
    structure \\
    Category analysis & 5 & Root cause names specific model component; cites $\geq$2
    session patterns; recommendations are engineer-implementable; severity matches
    session count; all fields present \\
    Verification & 3 & Catches factual inconsistencies; preserves correct claims
    unchanged; flags contradictory recommendations across categories \\
    Synthesis & 4 & Numbers match input data; priority justified with reasoning;
    $\geq$1 cross-cutting observation spans 2+ categories; no hallucinated categories \\
    Code suggestions & 4 & All file paths exist in codebase; targets root cause from
    analysis; implementable as described; no destructive changes \\
    \midrule
    Total & 25 & \\
    \bottomrule
  \end{tabular}
\end{table}

\begin{table}
  \caption{End-to-end cost breakdown for production diagnostic runs. Per-session judge
  costs are estimated from observed per-session rates applied to the full session
  population (GOOD + BAD). Code-suggestion generation cost is estimated per run from the
  model-code input context and the generated patches (Opus~4.6 rates), not from telemetry.}
  \label{tab:cost}
  \begin{tabular}{llrrrr}
    \toprule
    & & \multicolumn{2}{c}{Platform A} & \multicolumn{2}{c}{Platform B} \\
    & & \multicolumn{2}{c}{(96,801 sess.)} & \multicolumn{2}{c}{(101,594 sess.)} \\
    Stage & Model & Cost & \% & Cost & \% \\
    \midrule
    Upstream: Session judges & Flash & \$320.98 & 91.9 & \$250.21 & 99.0 \\
    Classification & Flash & \$26.29 & 7.5 & \$0.65 & 0.3 \\
    Consolidation & Flash & \$0.03 & 0.0 & \$0.00 & 0.0 \\
    Category analysis & Pro & \$0.49 & 0.1 & \$0.17 & 0.1 \\
    Verification & Pro & \$0.08 & 0.0 & \$0.05 & 0.0 \\
    Synthesis & Pro & \$0.03 & 0.0 & \$0.01 & 0.0 \\
    Code suggestions & Opus 4.6 & \$1.36 & 0.4 & \$1.76 & 0.7 \\
    \midrule
    AURA subtotal & & \$28.28 & 8.1 & \$2.64 & 1.0 \\
    Total (end-to-end) & & \$349.26 & & \$252.85 & \\
    \bottomrule
  \end{tabular}
\end{table}

\subsection{Cross-Platform Generalizability}
\label{sec:generalizability}

We ran AURA on production data from two streaming platforms with independent data
schemas, content catalogs, recommendation models, and user populations, asking whether
the same architecture produces meaningful findings on both without platform-specific
tuning. The core pipeline code is identical across platforms; platform-specific
configuration supplies data mappings, SQL fragments, prompt templates, taxonomies, and
significance thresholds. Despite independent taxonomy development, both platforms
converged on a closed 13-category enumeration sharing common concepts while differing in
platform-specific categories (\S\ref{sec:diagnostic-quality}).

\paragraph{From streaming to e-commerce.} The same portability argument extends across
domains, because the architecture never touches domain semantics. The pipeline consumes
sessions, a judge verdict, and a codebase, and every domain-specific element (data
mappings, taxonomies, prompts, thresholds) arrives through the same configuration layer
that ported Platform~B to Platform~A with no core-code changes.
Table~\ref{tab:ecommerce-mapping} makes the correspondence concrete. Each element of the
deployed streaming instantiation has a direct conceptual e-commerce analog, so an e-commerce tenant
can potentially onboard the way Platform~A did, through configuration.

\begin{table}
  \caption{The deployed streaming instantiation and its e-commerce analog. Onboarding is
  designed to be a configuration change.}
  \label{tab:ecommerce-mapping}
  \begin{tabular}{p{0.44\linewidth}p{0.46\linewidth}}
    \toprule
    Streaming (deployed) & E-commerce analog \\
    \midrule
    Session: impressions, clicks, watch time & Session: impressions, add-to-cart, purchase, return signals \\
    GOOD/BAD judge over engagement evidence & GOOD/BAD judge over conversion and satisfaction evidence \\
    Genre / sub-genre preference mismatch & Category and price-affinity mismatch \\
    Franchise affinity mismatch & Brand affinity mismatch \\
    Kid-profile safety (age-inappropriate content) & Age-restricted item compliance \\
    Post-launch content staleness & Out-of-stock and seasonal staleness \\
    Regional preference mismatch & Regional assortment and shipping mismatch \\
    Cross-shelf deduplication failure & Duplicate listings across carousels \\
    Ranker code, features, training pipeline & Ranker code, features, training pipeline (unchanged) \\
    \bottomrule
  \end{tabular}
\end{table}

\section{Deployment and Lessons}
\label{sec:deployment}

We are integrating AURA into the ML engineering workflow alongside existing A/B testing and metrics dashboards. The port from Platform B to Platform A required no core-pipeline changes (\S\ref{sec:generalizability}). Database-backed prompt templates support versioning and rollback, enabling prompt iteration without redeployment, while engineers review findings through a shared web interface. Model-version changes can affect prompt formatting and baseline distributions, so we treat prompts as versioned artifacts and re-evaluate them against historical baselines before upgrades.

\label{sec:safeguards}
LLM agents with access to production data and code need explicit safeguards at every
boundary. Every agent runs read-only against analytical data lakes, and the
code-implementation stage operates on a cloned, sandboxed repository. A named human owns
every change that ships. AURA generates pull requests, and every finding carries the raw session evidence (session IDs, severity, root-cause hypotheses) the engineer validates independently. Because a model-suggested
change that breaks production lands on a named owner, engineer review is a safety
mechanism rather than a convenience layer, and dismissal is a first-class outcome.

\label{sec:lessons}
Beyond the diagnostic results of \S\ref{sec:diagnostic-quality}, four operational choices
proved decisive for adoption. We closed the category vocabulary after an initial
discovery phase, and the taxonomy stopped drifting across runs. Engineers trusted findings
precisely because they could check them, so LLM-authored prose became packaging rather
than the deliverable. Per-step cost went on the screen instead of in a billing dashboard,
which made each run easy to justify. And the mandatory platform parameter that looked
redundant caught cross-platform data-mixing bugs at development time and let Platform~A
onboard with almost no new code.

The dominant lesson, and the hardest, was that every LLM output boundary needs a defensive
layer, implemented as the cross-stage validation layer of \S\ref{sec:validation}. Each
failure below surfaced in a baseline run and is now caught programmatically:
\begin{itemize}
  \item \textbf{Fabricated identifiers.} On a baseline Platform~B run (prior to the fixes
  in \S\ref{sec:diagnostic-quality}), the category-analysis stage populated session-ID
  fields with 904 fabricated UUIDs dressed up with real collection names to look
  plausible (e.g., \texttt{37a9...5147::Top 15 Today}). Post-verification sanitization
  now strips any ID absent from the input.
  \item \textbf{Severity miscalibration.} Over-inflated severity scores traced to the LLM
  inferring session counts. Passing counts and percentages into the prompt directly
  resolved it on both platforms simultaneously.
  \item \textbf{Schema and reference violations.} Malformed JSON is caught by defensive
  validation with a regex fallback. Code suggestions that referenced nonexistent files
  are now validated against the actual codebase.
  \item \textbf{Taxonomy coordination failures.} Closed taxonomies are brittle. For instance, when
  Platform~A's category database was wiped, restrictive prompt instructions collapsed all
  sessions into two categories until the taxonomy was re-embedded.
\end{itemize}

\section{Future Work}

Two directions define our next steps: stronger evidence and a closed loop. On evidence, a
held-out human audit of diagnostic quality and comparisons against external baselines
remain for future work. On closing the loop, candidate training and A/B testing remain manual behind the
engineer checkpoint. Automating them so discovered failure categories feed back into
model improvements is the natural extension of our closed-loop validation. In
parallel, we are scaling agent orchestration and cross-run memory so that each run reuses
what earlier runs discovered and avoids re-diagnosing known failures.

We further plan to give the judges richer context, including agentic tools, knowledge-graph
representations, and global and local engagement signals, and to expand from a single judge
to a committee that reduces the biases of any one LLM. Finally, we aim to develop adaptive
judges that condition on persona, with success criteria tuned per ranker and retriever.

\section{Conclusion}

Aggregate metrics tell recommender teams whether a model is improving on average, not
what is failing or for whom. That knowledge is a gap in the existing process that recent AI
agentic technology can help close. AURA closes that gap. It reads thousands to millions of
production sessions and returns a ranked taxonomy of the ways the system fails real
users. 
Then it reads the recommender's own code and says what to change, and why.

Run on production data from both platforms, AURA
produces platform-specific taxonomies, runs at low end-to-end cost, and surfaces findings
engineers can check against raw evidence. Two lessons stood out: locking the
category vocabulary after initial discovery beat open-ended classification, and engineer
trust came from structured evidence, not LLM authority. Offline closed-loop validation
adds a cheap check that rejects fixes before they reach a live experiment. In its first
application, the most useful thing AURA did was refuse to flatter us: the two
genre-targeted fixes an engineer would have tried first did not resolve the diagnosed
failure, and learning that took an offline screen rather than an A/B test.

\bibliography{references}

\end{document}